# Heptagonal Photonic Crystal Fiber for Dispersion Compensation with a Very Low Confinement Loss

Md Borhan Mia, Mohammad Faisal, Syeda Iffat Naz, Kanan Roy Chowdhury, and Animesh Bala Ani

*Abstract*— This paper presents a photonic crystal fiber (PCF) with heptagonal core and heptagonal cladding for dispersion compensation. Different optical properties of the suggested PCF are explored using the finite element method. The proposed dispersion compensating PCF (DC-PCF) exhibits a very large negative chromatic dispersion of −940 ps/(nm-km) at 1550 nm wavelength. The dispersion variation of the DC-PCF is −420.1 to −1160 ps/(nm-km) in the wavelength range of 1390 to 1700 nm (310 nm band). The relative dispersion slope is 0.0036 nm$^{-1}$ which is a perfect match with the standard single mode fibers. Moreover, it exhibits a very low confinement loss of about 10$^{-5}$ dB/km and low nonlinearity of 45 W$^{-1}$km$^{-1}$ at 1550 nm wavelength. Since the suggested DC-PCF has very high negative dispersion and low nonlinearity, it could be a potential candidate for broadband dispersion compensation in fiber-optic communication.

*Index Terms*—Confinement loss; Chromatic dispersion; Finite element method; Photonic crystal fiber

## I. INTRODUCTION

Wavelength division multiplexed (WDM) and dense wavelength division multiplexed (DWDM) based fiber-optic transmission networks are widely used to transport high bit rate and ultra-high bit rate data [1]. Due to growing demand of various services, e.g., wireless digital cameras, video conferencing, online video streaming, gaming on demand, IPTV, telemetry, cloud computing, Internet of things (IoT) etc., the data over the telecommunication links are increasing day by day. In WDM systems, the bit rate of 40 Gb/s has been widely used [2]. Moreover, data rate per channel of 100 Gb/s [3] and 400 Gb/s [4] have been realized in order to transport massive data over WDM and DWDM systems for long-haul transmission. The optical fiber has mainly two linear impairments, namely, attenuation and dispersion which require to be mitigated. Between the two, the chromatic dispersion is more detrimental and it imposes a considerable limitation as it broadens the optical pulses propagating through the fiber. With the increase of demand for capacity, the bit rate increases and the pulse width decreases which consequently further enhances the pulse broadening and drastically reduces the performance of the systems. Therefore, dispersion compensation is indispensable in fiber-optic communication.

Standard single mode fibers (SMFs) are mostly deployed in fiber-optic transmission links which have a typical dispersion of 15 to 20 ps/(nm-km) at 1550 nm wavelength. This will cause a huge accumulated dispersion along the transmission line. To compensate this dispersion of the SMFs, different strategies have been adopted such as conventional dispersion compensating fibers (DCFs) [5], electronic dispersion compensation [6], Bragg grating fibers [7] and optical phase conjugation [8] etc. Among these techniques, DCFs are widely used due to their negative dispersion, low loss and low nonlinearity [9]. Furthermore, compensation is done in optical domain with ease of coupling with SMF, if we use DCFs. Dispersion compensating PCF (DC-PCF) offers additional benefit by tailoring the optical properties like dispersion, loss and birefringence etc. We use DC-PCF to achieve desirable negative dispersion with suitable slope, low or high birefringence, and further reduction in the losses. Therefore, DC-PCFs are preferred over DCFs.

In recent years, photonic crystal fibers (PCFs) or micro-structure optical fibers (MOFs) consisting of air hole channel running down their length have earned a lot of interest due to their sui-generis properties and potential applications. Their optical properties are easily engineered since they have various tuneable design parameters such as pitch, diameter, shape and number of air holes and air-hole rings around the core as well as the cladding region [5]–[10]. These flexibilities of the PCFs make it suitable and potential to supersede SMFs and DCFs. There are some other attractive properties that can be tailored from the PCFs such as improved birefringence, nonlinearity, large effective area, splice loss, bending loss etc. Physical shapes of the PCFs are quite feasible today because of the invention of fabrication techniques like stack and draw [11], drilling and extrusion [12], sol-gel casting [13] etc.

The idea to use PCF in dispersion compensation was first introduced by Birks et al. [14] who proposed a PCF showing chromatic dispersion of $-2\times10^{-3}$ ps/(nm-km) at 1550 nm wavelength. Kim et al. [15] has proposed a PCF displaying dispersion of $-156 \pm 0.5$ ps/(nm-km) over the wavelength range of C+L bands. However, the core and cladding consist of elliptical air holes which are difficult to fabricate, difficult to coupling and dispersion value is very low. Habib et al. [16] proposed a PCF of dispersion $-588$ ps/(nm-km) at 1550 nm. However, the confinement loss is $10^{-1}$ dB/m which is very high. In [17], Haque et al. proposed another design with chromatic dispersion of $-790.12$ ps/(nm-km) and the confinement loss of $10^{-4}$ dB/km at 1550 nm wavelength. However, dispersion coefficient varies from $-248$ to $-1069$ ps/(nm-km) over E to L bands. In another design, Habib et al. [18] proposed an MOF for which dispersion coefficient is $-130$ to $-360$ ps/(nm-km) in the wavelength range of 1400 nm to 1500 nm (100 nm band). Moreover, the background material used here is not silica but organic materials like propanol, butanol and ethanol. In [19], Hasan et al. proposed a PCF of dispersion $-555.93$ ps/(nm-km) at the wavelength of 1550 nm. Besides, dispersion co-efficient varies from $-388.72$ to $-723.1$ ps/(nm-km) ranging from 1460 to 1625 nm wavelength. Additionally, confinement loss is also high, of the order of $10^{-2}$ dB/km at 1550 nm wavelength. In [20], a structure using 5 rings of air holes is proposed whose dispersion varies from $-386.57$ to $-971.44$ ps/(nm-km) over the wavelength ranging from 1400 nm to 1610 nm. The chromatic dispersion at 1550 nm is $-790.12$ ps/(nm-km). However, the design is much complex because of air-holes with different diameters which are angularly rotated and different pitches. All these make the design a fabrication challenge. In [21], the obtained average dispersion is $-138$ ps/(nm-km) from 980 nm to 1580 nm, however, again the design is complex since the core region consists of pentagonal shaped air holes. Besides, the confinement loss is also high. In [22], Habib et al. suggested a single-mode fiber of dispersion $-712$ ps/(nm-km). Additionally, dispersion varies from $-300$ to $-1000$ ps/(nm-km) in the wavelength range of 1340 nm to 1640 nm. Monfared et al. [23] investigated a PCF and obtained dispersion of $-2108$ ps/(nm-km) at 1550 nm wavelength. However, necessary information regarding relative dispersion slope (RDS) is missing. Furthermore, core is elliptical shape doped with $GeO_2$ which makes it costlier and suffered fabrication difficulty. In [24], the chromatic dispersion at 1550 nm wavelength is $-672$ ps/(nm-km). However, there are two elliptical air holes in the core and the confinement loss is very high. Additionally, dispersion varies from $-470$ to $-850$ ps/(nm-km) over S to L bands. The PCF in [25] exhibits chromatic dispersion of $-1054.4$ ps/(nm-km) at 1550 nm wavelength and RDS value match with SMFs. Notwithstanding, the structure is hybrid which increases the complexity and there is no information of confinement loss. Besides, the dispersion varies from $-270$ to $-1100$ ps/(nm-km) in the wavelength range of 1340 to 1640 nm (300 nm bands). The large negative dispersion is shown in [26], [27], however, information regarding either RDS or confinement loss is missing and the core is doped in [26] and dispersion is fluctuating in [27].

In this paper, we propose a PCF with heptagonal core and cladding for dispersion compensation. The proposed DC-PCF shows a negative dispersion of $-940$ ps/(nm-km) at the wavelength of 1550 nm which is higher than that of recently reported literatures [16-20]. The dispersion variation is of $-420.1$ to $-1160$ ps/(nm-km) in the wavelength range of 1390 to 1700 nm (310 nm bands). The RDS value of the proposed DC-PCF is 0.0036 nm$^{-1}$, which is perfectly matched with the SMFs. Besides, the confinement loss is very low in the order of $10^{-5}$ dB/km. The proposed DC-PCF operates on single mode since $V_{eff}$ parameter is less than π from wavelength of 1390 nm to 1700 nm (310 nm bands).

## II. GEOMETRY OF THE DESIGN

The cross sectional view of the proposed DC-PCF is demonstrated in Fig. 1. Design is simple since the core and cladding consist of circular air holes in a shape of heptagon. For the background material silica is used, which is industrially available. The proposed design consists of total nine air hole rings divided into two sectors, i.e. inner core region and outer cladding. Inner and outer cladding both have the heptagonal shape. The core region is comprised of three air-hole rings with the pitch value of $\Lambda_1 = 0.781$ μm and the diameter of each hole is $d_1 = 0.631$ μm for the optimum case. It is evident that change of diameter of air hole or pitch near the core region strongly affects the dispersion property of the fiber. We check the effect of both in core as well as cladding for our work. The angle between two adjacent holes in core is 51.43°. For the outer cladding, there are 6 air-hole rings with the pitch value of $\Lambda_2 = 0.867$ μm and the hole diameter of $d_2 = 0.80$ μm for the optimum results. As seen from the Fig. 1, the core has identical air-hole with diameter of $d_1$ and the cladding with diameter of $d_2$. The air filling fraction in core and the cladding are chosen to be $d_1/\Lambda_1=0.81$ and $d_2/\Lambda_2=0.92$, respectively to achieve optimum result.

## III. MODEL AND SIMULATION METHOD

To investigate the optical properties of the designed fiber, finite element method (FEM) is used. COMSOL MULTIPHYSICS software, version 5.0 is used as the design simulator. A circular perfectly matched layer (PML) is positioned outside the outermost ring to model the leakage and it produces no reflection.

Maxwell's equations are solved in layers comprising a finite number of homogeneous isotropic regions ended by PML. By solving the eigenvalue problem of the Maxwell's curl equation, the effective refractive index can be obtained. Maxwell's equation [28] is expressed as

$$\vec{\Delta} \times \left\{ \frac{1}{[s]} \cdot (\vec{\Delta} \times \vec{E}) - k_0^2 \eta^2 [s] \vec{E} \right\} = 0 \qquad (1)$$

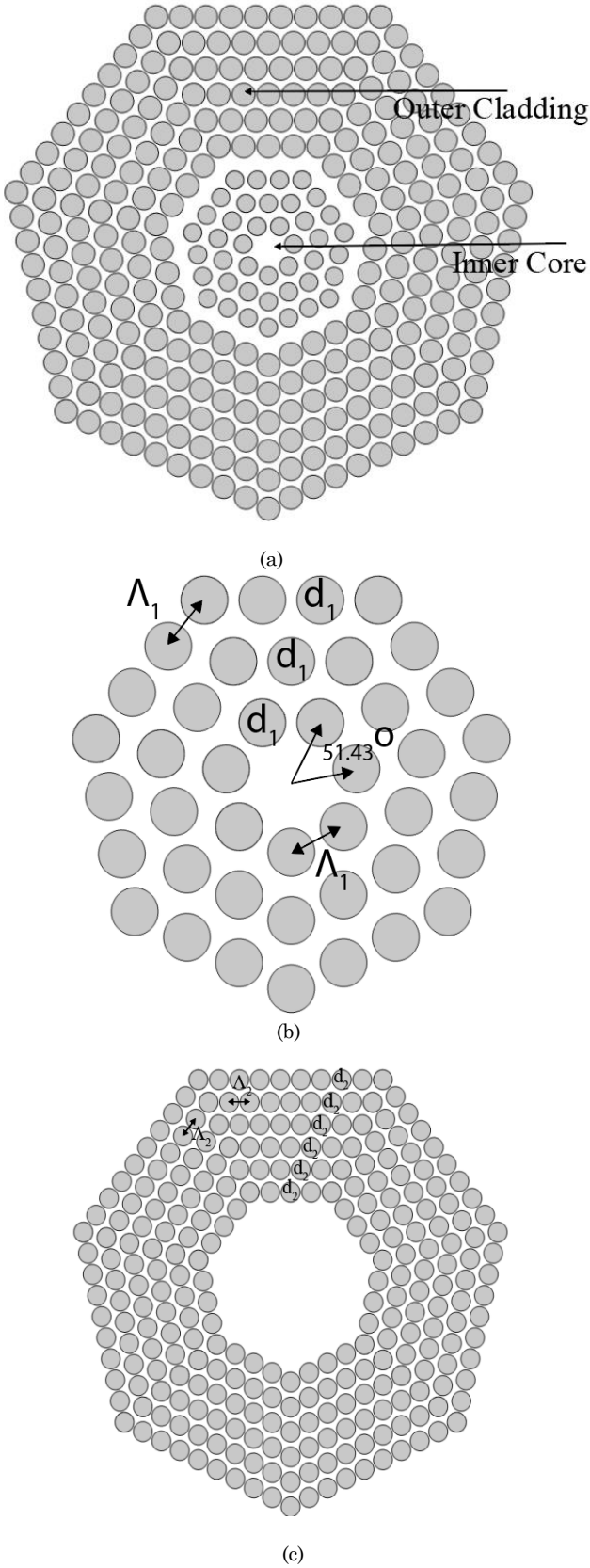

Fig. 1. Transverse cross section of the suggested DC-PCF with the optimum parameters $d_1$, $d_2$, $\Lambda_1$ and $\Lambda_2$. (a) Full cross sectional view of DC-PCF, Cross sectional view of (b) magnified core (c) Cladding

where $E$ is the electric field vector, $k_0 = \frac{2\pi}{\lambda}$ is the free space wave number, $\eta$ is the refractive index, [s] is the matrix of the PML circular layers and $[s]^{-1}$ is the inverse matrix. Once the modal refractive index $\eta_{eff}$ is obtained, other parameters like chromatic dispersion $D(\lambda)$, nonlinear coefficient $\gamma$, confinement loss $L_c$, effective area $A_{eff}$, birefringence $B$, and $V$ parameter etc. can be obtained from their respective equation [29]–[32].

$$D(\lambda) = -\frac{\lambda}{c}\frac{d^2 \operatorname{Re}[\eta_{eff}]}{d\lambda^2} \quad (2)$$

$$B = |\eta_{eff}^x - \eta_{eff}^y| \quad (3)$$

$$A_{eff} = \frac{\left(\iint |E|^2 \, dxdy\right)^2}{\iint |E|^4 \, dxdy} \quad (4)$$

$$L_c = \frac{20 \times 10^6}{\ln(10)} k_0 \operatorname{Im}[\eta_{eff}] \quad (5)$$

where, $\operatorname{Re}[\eta_{eff}]$ and $\operatorname{Im}[\eta_{eff}]$ are the real and the imaginary part of the effective refractive indices, respectively; $\eta_{eff}^x$ and $\eta_{eff}^y$ are the refractive indices of the x and y polarization mode, respectively, $E$ is the electric field vector, $k_0$ is the free space wave number, $\lambda$ is the operating wavelength and c is the velocity of the light in vacuum. The material dispersion is considered by applying Sellmeier formula in simulation

The change of dispersion with a small change in wavelength is known as dispersion slope and it is necessary to calculate the slope mismatch with other fibers. Dispersion slope is calculated by the following equation [33].

$$S(\lambda) = \frac{dD(\lambda)}{d\lambda} \quad (6)$$

The ratio of dispersion slope to dispersion is defined as relative dispersion slope (RDS). The RDS value of the proposed fiber is calculated by using the following equation [34].

$$RDS = \frac{S_{SMF}(\lambda)}{D_{SMF}(\lambda)} = \frac{S_{DC-PCF}(\lambda)}{D_{DC-PCF}(\lambda)} \quad (7)$$

where, $S_{SMF}(\lambda)$ and $D_{SMF}(\lambda)$ are the dispersion slope and dispersion for the SMF, respectively. Similarly, $S_{DC-PCF}(\lambda)$ and $D_{DC-PCF}(\lambda)$ are for the dispersion compensating fibers. Once we obtain the RDS of the DC-PCFs close to that of the standard single mode fibers (SMFs), our target will be accomplished for the broadband dispersion compensation in fiber-optic communication.

The mode property of the proposed DC-PCF is scrutinized carefully. There is a cut off frequency for each mode below which the mode cannot propagate. The normalized frequency or V-parameter of the optical fiber is a very important quantity which is used to determine the propagating modes. If V-parameter of the fiber is less than $\pi$, fiber will allow only one mode and effectively acts as a single-mode fiber. The V-parameter, $V_{eff}$ is expressed by the following equation [35].

$$V_{eff} = \frac{2\pi\Lambda}{\lambda}\sqrt{\eta_{eff}^2 - \eta_{FSM}^2} \qquad (8)$$

where, $\eta_{eff}$ and $\eta_{FSM}$ are the refractive indices of the fundamental mode and fundamental space filling mode, respectively.

## IV. RESULTS AND DISCUSSION

Fundamental mode field distribution and the effective refractive index of the proposed DC-PCF are presented in Fig. 2 (a) and (b), respectively. Light is well confined in the core region since higher refractive index is achieved in the core region than the cladding which results in high negative value of dispersion.

It is seen form the Fig. 2 (b) that the effective refractive index is 1.28 at 1550 nm wavelength. With the increase of wavelength, refractive index decreases indicating that power is well confined in core region at lower wavelength.

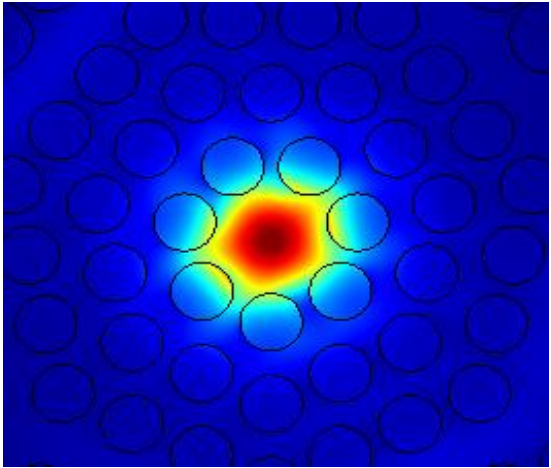

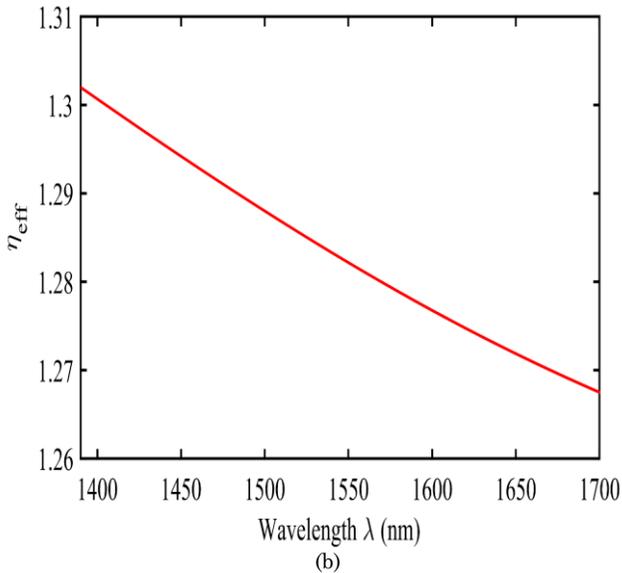

Fig. 2. With the parameter $d_1 = 0.631$ µm, $\Lambda_1 = 0.781$ µm, $d_2 = 0.80$ µm and $\Lambda_2 = 0.867$ µm for the proposed DC-PCF, (a) Poynting vector profile and (b) Refractive index vs. wavelength

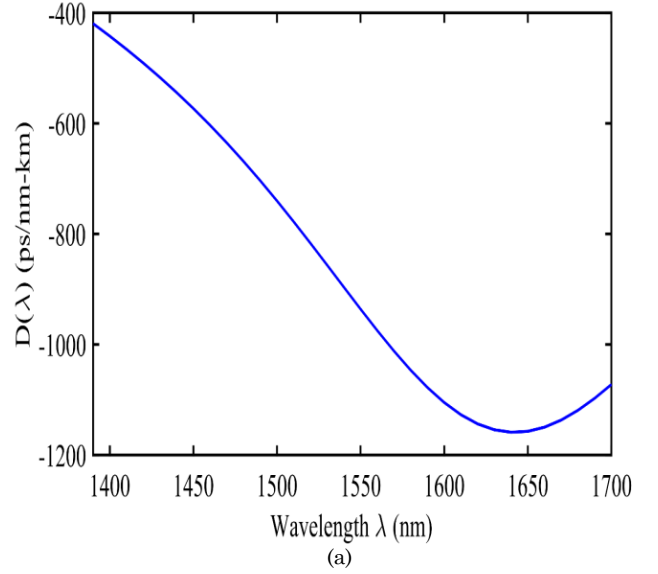

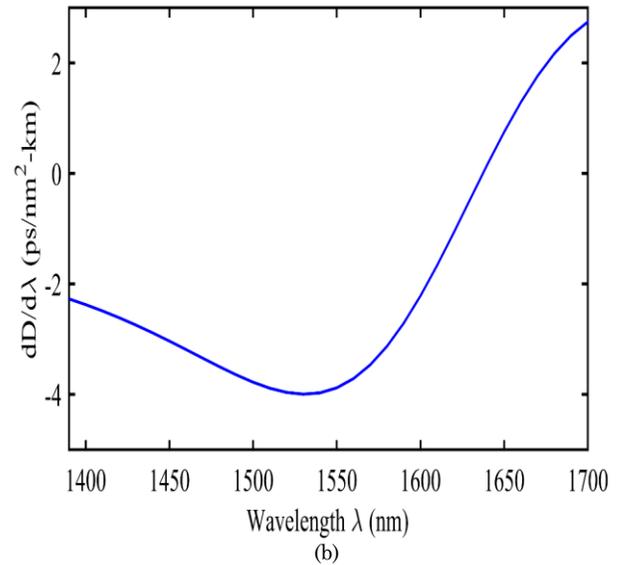

Fig. 3. Optimum parameter $d_1 = 0.631$ µm, $\Lambda_1 = 0.781$ µm, $d_2 = 0.80$ µm and $\Lambda_2 = 0.867$ (a) Chromatic dispersion (b) dispersion slope as a function of wavelength.

Firstly, we have changed core diameter $d_1$ keeping other parameters constant, i.e., $\Lambda_1 = 0.781$ µm, $\Lambda_2 = 0.867$ µm and $d_2 = 0.80$ µm. Taking the value of $d_1$ as 0.621, 0.631, 0.641 µm, the corresponding chromatic dispersions are found −890, −940, −1000 ps/(nm-km), respectively. Calculated

RDS values are 0.00416, 0.0036 and 0.0049 nm$^{-1}$ at $d_1$ = 0.621, 0.631 and 0.641 µm, respectively at 1550 nm wavelength. The effect of variation of diameter $d_1$ on chromatic dispersion is depicted in Fig. 4. Moreover, the confinement loss of the proposed PCF to the corresponding diameters, $d_1$ = 0.621, 0.631, 0.641 µm are 4.64×10$^{-6}$, 1×10$^{-5}$ and 1.46×10$^{-5}$ dB/km, respectively and it is seen in Fig. 5.

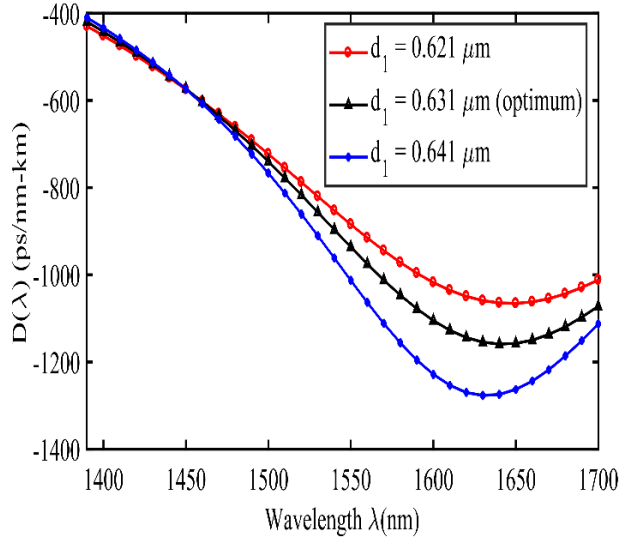

Fig. 4. Chromatic dispersion of the signified DC-PCF as function of wavelength with the parameter $d_1$ taken as 0.621 µm, 0.631 µm and 0.641 µm keeping $d_2$, $\Lambda_2$ and $\Lambda_1$ unaltered.

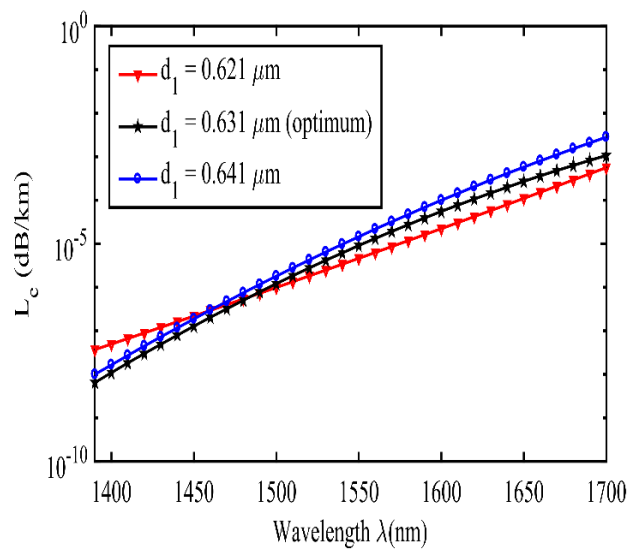

Fig. 5. Confinement loss vs. wavelength with parameter $d_1$ taken as 0.621 µm, 0.631 µm and 0.641 µm keeping $d_2$, $\Lambda_2$ and $\Lambda_1$ unchanged.

Later on, we keep $d_1$ = 0.631 µm, $d_2$ = 0.80 µm and $\Lambda_2$ = 0.867 µm unchanged and vary the value of $\Lambda_1$ in core. The chromatic dispersion of the proposed PCF with the changing values of $\Lambda_1$ is shown in the Fig. 6. $\Lambda_1$ is chosen as 0.779, 0.781 and 0.782 µm while the calculated dispersion is −840, −940 and −1010 ps/(nm-km), respectively. The calculated RDS values are 0.0040, 0.0036, 0.0043 nm$^{-1}$, respectively at 1550 nm wavelength. The confinement loss for varying core pitch, $\Lambda_1$ is demonstrated in the Fig. 7. It is seen from the Fig. 7 the confinement loss is 6.14×10$^{-6}$, 1×10$^{-5}$ and 1.28×10$^{-5}$ dB/km for the pitch of $\Lambda_1$ = 0.779, 0.781 and 0.782 µm, respectively.

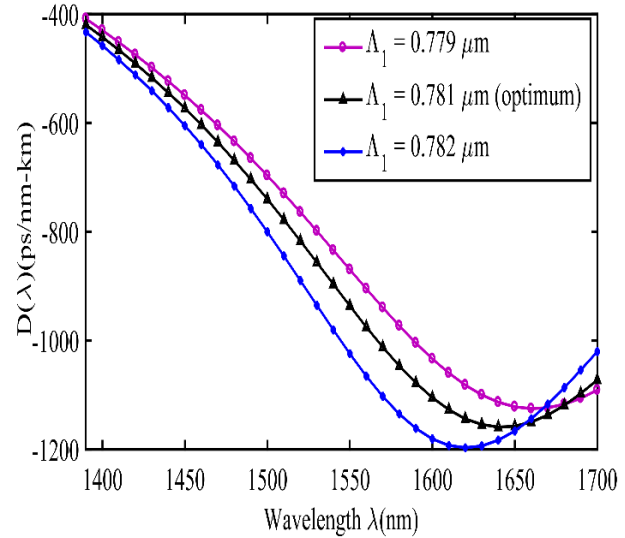

Fig. 6. Chromatic dispersion of the denoted DC-PCF as function of wavelength with the parameter $\Lambda_1$ taken as 0.779 µm, 0.781 µm and 0.782 µm keeping $d_2$, $\Lambda_2$ and $d_1$ unchanged.

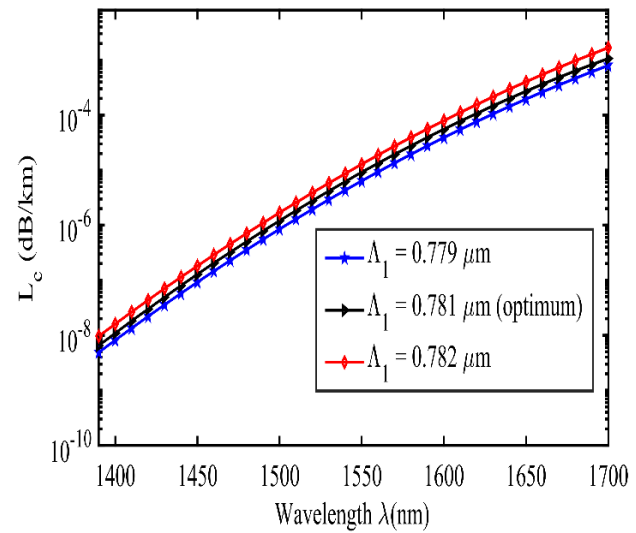

Fig. 7. Confinement Loss of the proposed DC-PCF as function of wavelength with the parameter $\Lambda_1$ taken as 0.779 µm, 0.781 µm and 0.782 µm keeping $d_2$, $\Lambda_2$ and $d_1$ untouched.

The above discussion reveals that whether $d_1$ value is increased or decreased, either way, RDS value moves away from 0.0036nm$^{-1}$ though the chromatic dispersion increases with the increment of $d_1$ at the wavelength of 1550 nm. On the other hand, the increment or decrement of $\Lambda_1$ has the same effect as $d_1$. So the optimum core diameter and pitch are chosen to be $d_1$ = 0.631 µm and $\Lambda_1$ = 0.781 µm, respectively. The confinement loss from the Fig. 5 reveals that with the increment of $d_1$ loss will increase since optical field moves towards the cladding region for large $d_1$. From the Fig. 7, the confinement loss increases for the large pitch $\Lambda_1$, due to mode field travels to cladding region.

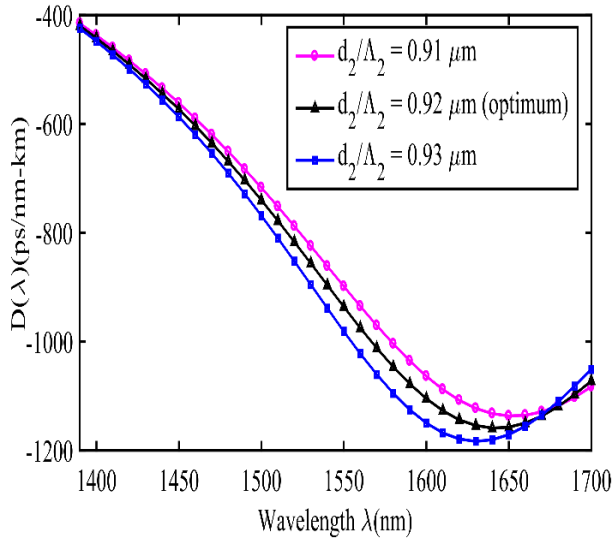

Fig. 8. The proposed DC-PCF's chromatic dispersion as a function of wavelength changing $d_2/\Lambda_2$ from 0.91 µm to 0.93 µm keeping other parameters unaltered.

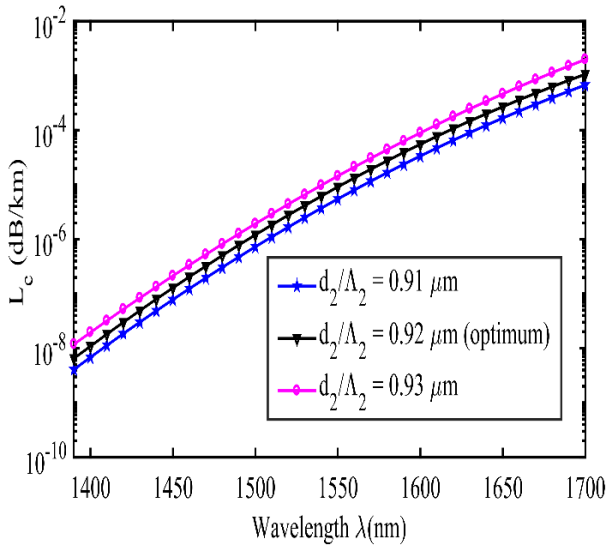

Fig. 9. Confinement loss vs. wavelength for the changing $d_2/\Lambda_2$ from 0.91 µm to 0.93 µm keeping other parameters unaltered.

To explore the effect of the outer cladding, we have changed the $d_2/\Lambda_2$ ratio keeping $d_1 = 0.631$ µm and $\Lambda_1 = 0.781$ µm unaltered. The value of $d_2/\Lambda_2$ is taken as 0.91, 0.92 and 0.93 µm and the corresponding dispersion curves are shown in the Fig. 8. Since the parameter $d_2/\Lambda_2$ is changing from 0.91 µm to 0.93 µm, the chromatic dispersion has increased. The chromatic dispersions are −900, −940 and −980 for $d_2/\Lambda_2$ = 0.91, 0.92 and 0.93, respectively at communication band. However, RDS values at 1550 nm are obtained as 0.004, 0.0036, and 0.0048 nm$^{-1}$, respectively. The confinement loss is demonstrated in the Fig. 9 and found $5.50×10^{-6}$, $1×10^{-5}$ and $1.50×10^{-5}$ dB/km, respectively at 1550 nm wavelength. The confinement loss is decreased below the value of $d_2/\Lambda_2 = 0.92$ and increased above the below of $d_2/\Lambda_2 = 0.92$. This is because the optical field is more confine when the ratio of $d_2/\Lambda_2 = 0.92$. Therefore, the optimum parameters of the proposed fiber are taken to be $d_2/\Lambda_2 = 0.92$, $d_1 = 0.631$ µm and $\Lambda_1 = 0.781$ µm.

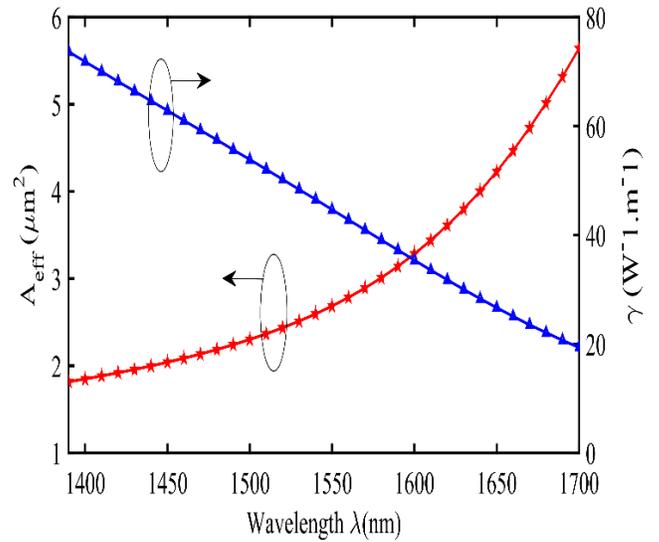

Fig. 10. Nonlinear co-efficient and effective area as a function of wavelength for the optimum parameter of the Proposed DC-PCF.

The effective area and the nonlinearity of the proposed fiber are depicted in Fig. 10 for the optimum values of parameters $d_1$, $d_2$, $\Lambda_1$ and $\Lambda_2$. It is clear from the Fig. 10 that at 1550 nm wavelength the effective area is 2.5 µm$^2$ and the nonlinear co-efficient is 45 W$^{-1}$km$^{-1}$ which is low. Therefore, the proposed fiber won't be much affected by the non-linearity.

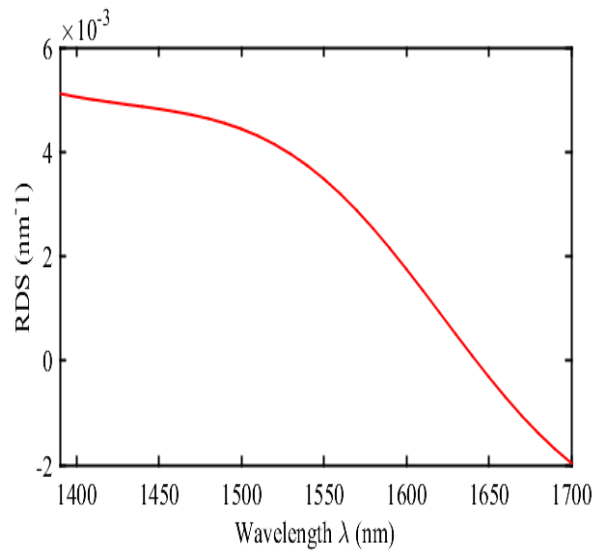

Fig. 11. RDS of the proposed DC-PCF vs. wavelength for the optimum parameters $d_1$, $\Lambda_1$, $d_2$ and $\Lambda_2$.

The RDS of the proposed DC-PCF is shown in the Fig. 11 and it is realized the RDS value of recommended DC-PCF is close to that of SMFs; nominally it is 0.0036 nm$^{-1}$ at 1550 nm wavelength. Since RDS is used to judge the dispersion compensation (DC), the proposed fiber is a perfect candidate for broadband dispersion compensation.

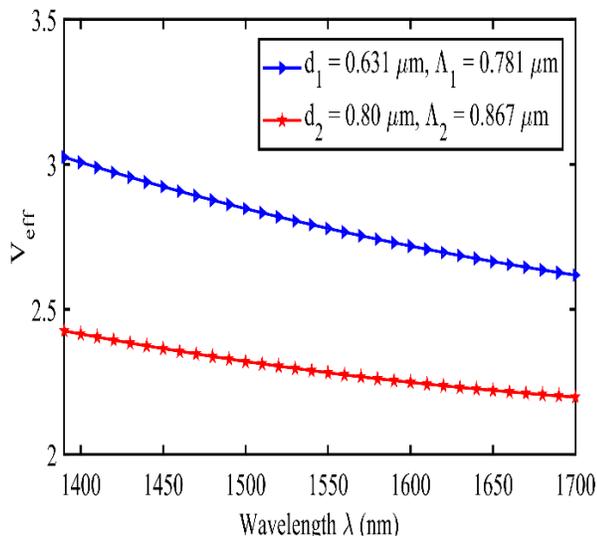

Fig. 12. V-parameter of the proposed DC-PCF as a function of wavelength for the inner and outer cladding with the optimum geometric entities.

The DC-PCF can be operated as a SMF if V-parameter, $V_{eff} \leq \pi$. If $V_{eff}$ is greater than the π, higher order modes get associated with it. The V-parameter of the proposed DC-PCF is presented in Fig. 12. From the Fig. 12, the obtained value for $V_{eff}$ is less than π from 1390 nm to 1760 nm wavelength (310 nm bands). So, it is ensured that the proposed DC-PCF will operate as single-mode over the O+E+S+C+L bands.

Fabrication viability is the major issue in realization of PCFs. Heterogeneous air hole diameters at the cladding region already has been fabricated using the stack-and-draw technique [36]. Bise et al. [13] used sol-gel technology to fabricate microstructure optical fibers with missing air holes. Even PCFs with high air filling fraction have been reported fabrication viability [37], [38]. All these methods could be used to fabricate our proposed DC-PCF, however, stack-and-draw technique will be more convenient.

Finally in Table 1, we compare our work with some recently published works for compensating dispersion. It is seen from the table, our proposed DC-PCF has offered much better results with simpler design.

TABLE I
COMPARISON OF PROPOSED DC-PCF WITH RECENT ARTICLES

| Ref | D at 1550 nm (ps/(nm-km)) | $L_c$ at 1550 nm (dB/km) | Dispersion variation (ps/(nm-km)) |
|---|---|---|---|
| 16 | −588 | $10^3$ | −400 to −725 (165 nm bands) |
| 17 | −790.12 | $10^{-4}$ | −248.65 to −1069 (270 nm bands) |
| 19 | −555.93 | × | −388.72 to −723.1 (165 nm bands) |
| 20 | −790.12 | $10^{-2}$ | −386.57 to −971.44 (210 nm bands) |
| 21 | −712 | × | −200 to −1000 (300 nm bands) |
| 22 | −672 | $10^{-1}$ | −470 to −850 (165 nm band) |
| Proposed DC-PCF | −940 | $10^{-5}$ | −420.1 to −1160 (310 nm band) |

## V. Conclusion

Heptagonal lattice structure DC-PCF is presented in this paper which exhibits a very high chromatic dispersion of −940 ps/(nm-km) at 1550 nm wavelength. The proposed DC-PCF shows dispersion variation from −420.1 to −1160 ps/(nm-km) covering the wavelength from 1440 nm to 1700 nm. To our best knowledge, this is highest result compared to recently published articles [16-17], [19-20], [22], [24]. The RDS of the suggested fiber is perfectly matched with standard SMF. The proposed fiber demonstrates a very low confinement loss of $10^{-5}$ dB/km at the wavelength of 1550 nm. In addition, the proposed DC-PCF operates on single mode over a wide telecom band. So, our suggested DC-PCF can be a suitable choice for broadband dispersion compensation.

## Acknowledgment

This article is based on our own research work and we have not received any fund from any organization.

Md Borhan Mia was with the Electrical and Electronic Engineering Department, Bangladesh University of Engineering and Technology, Dhaka, Bangladesh.

Dr. Mohammad Faisal is with the Electrical and Electronic Engineering Department, Bangladesh University of Engineering and Technology, Dhaka, Bangladesh.

Syeda Iffat Naz was with the Electrical and Electronic Engineering Department, Bangladesh University of Engineering and Technology, Dhaka, Bangladesh.

Kanan Roy Chowdhury was with the Electrical and Electronic Engineering Department, Chittagong University of Engineering and Technology, Chittagong, Bangladesh.

Animesh Bala was with the Electrical and Electronic Engineering Department, Bangladesh University of Engineering and Technology, Dhaka, Bangladesh.